\documentclass[twocolumn,showpacs,preprintnumbers,amsmath,amssymb]{revtex4}

\usepackage{graphicx}
\usepackage{dcolumn}
\usepackage{bm}

\newcommand{\be}{\begin{equation}}
\newcommand{\ee}{\end{equation}}
\def\nn{\nonumber}
\def\bea{\begin{eqnarray}}
\def\eea{\end{eqnarray}}

\newcommand{\eq}[1]{(\ref{#1})}

\def\d{\delta}

\def\g{\gamma}

 \def\L{\Lambda} 
\def\m{\mu}

\def\cM{{\cal M}}  
  \def\cR{{\cal R}}

\def\R{{\mathbb R}}

\def\one{\mbox{1 \kern-.59em {\rm l}}}

\def\({\left(}
\def\){\right)}
\def\diag{\mbox{diag}}
\def\Tr{{\rm Tr}}

\begin{document}

\preprint{UWTHPh-2009-02}

\title{Cosmological solutions of emergent 
noncommutative gravity}

\author{Daniela Klammer}
 \email{daniela.klammer@univie.ac.at}
\author{Harold Steinacker}%
 \email{harold.steinacker@univie.ac.at}
\affiliation{Fakult\"at f\"ur Physik, Universit\"at Wien.} 

\date{\today}

\begin{abstract}

Matrix models of Yang-Mills type lead to an emergent gravity theory,
which may not require fine-tuning of a cosmological constant.
We find cosmological solutions of Friedmann-Robertson-Walker type. 
They generically have a big bounce, and an early inflation-like phase 
with graceful exit. The  mechanism is purely 
geometrical, no ad-hoc scalar fields are introduced. 
The solutions are stabilized 
through vacuum fluctuations and are thus compatible 
with quantum mechanics.
This leads to a Milne-like universe after inflation,
which appears to be in remarkably good agreement with observation
and may provide an alternative to standard cosmology.

\end{abstract}


\maketitle

Quantum field theory and general relativity provide
the basis of our present understanding 
of fundamental forces and matter. However, there 
is up to now no satisfactory way to reconcile them
in a consistent quantum theory.
General arguments based on quantum mechanics and 
general relativity suggest  a ``foam-like'' or quantum
structure at the Planck scale $10^{-33}$ cm. 
This problem has its most dramatic manifestation 
in the cosmological constant problem: 
the small but non-vanishing cosmological constant
in the currently accepted $\Lambda$CDM model 
is in striking contradiction with quantum mechanical expectations, 
which are off by at least $60$ orders of magnitude.
No satisfactory solution of this problem within the 
conventional frameworks 
has been found. 

We point out in this letter that 
emergent gravity on non-commutative (NC) spaces
may provide a resolution of these problems. 
The starting point are matrix models of Yang-Mills type
\be
S_{YM} = - \Tr [X^a,X^b] [X^{a'},X^{b'}] 
\eta_{aa'}\eta_{bb'} ,
\label{YM-action-extra}
\ee
supplemented by suitable fermionic terms.
Such models have been discussed in the context of NC gauge theory
and string theory. 
Here $X^a,  \,\, a=1,..., D$ 
are (infinite-dimensional) hermitian matrices,
and $\eta_{ab} = {\rm diag}(1,1,..., \pm 1)$ 
in the Euclidean resp.  Minkowski case.
The basic hypothesis of our approach is that 
space-time is realized as 3+1-dimensional 
NC ``brane'' solution of such a model. 
The effective geometry and gravity on  
such a brane was clarified recently 
\cite{Steinacker:2007dq,Steinacker:2008ya,Steinacker:2008ri}:
the effective metric is not fundamental but 
depends on the embedding and the Poisson structure, 
reminiscent of the open string metric \cite{Seiberg:1999vs}. 
This leads to an emergent gravity closely related to NC gauge theory, 
as anticipated in \cite{Yang:2006mn,Rivelles:2002ez}.
Among these models, the IKKT model \cite{Ishibashi:1996xs} 
is singled out by maximal supersymmetry,  required by
consistency at the quantum level. 
This implies in particular $D=10$, 
which we assume from now on.
This model is a candidate for a theory of 
all fundamental interactions and matter.

The model admits $4$-dimensional 
noncommutative spaces $\cM_\theta \subset \R^{{10}}$ as 
solution, interpreted as space-time embedded in ${10}$ dimensions. 
This can be seen by splitting the matrices as 
\be
X^a = (X^\mu,\phi^i), \qquad \mu = 1,...,4, 
\,\,\, i=1, ..., 6 
\label{extradim-splitting}
\ee
where the ``scalar fields'' $\phi^i=\phi^i(X^\mu)$ are assumed to 
be functions of $X^\mu$. 
We only consider the {\em semi-classical}  
limit of such a quantum space, indicated by $\sim$.
Then $X^\mu \sim x^\mu$ is interpreted as quantization of 
coordinate functions on $\cM$, 
$\phi^i(x)$ define the embedding of a 4-dimensional 
submanifold $\cM \subset \R^{10}$, and
\be 
[X^\mu,X^\nu] \sim i\theta^{\mu\nu}(x), \quad  \mu,\nu = 1,..., 4
\label{theta-induced}
\ee 
can be interpreted as Poisson structure on $\cM$.
The physical fields arise from fluctuations in the matrix model
around such a background. Therefore they 
live only on the brane $\cM$, and there is no higher-dimensional ``bulk''
which could carry any propagating degrees of freedom,
unlike in string theory or standard braneworld-scenarios.
As shown in \cite{Steinacker:2008ya,Steinacker:2008ri}, 
the effective metric for all scalar, gauge and fermionic
fields propagating on $\cM$ is given by 
\bea  
G^{\mu\nu}(x) &=& e^{-\sigma}\, \theta^{\mu\mu'}(x) \theta^{\nu\nu'}(x) 
 g_{\mu'\nu'}(x), \label{G-def-general}  \\
e^{-\sigma} &=& \rho\, |g_{\mu\nu}|^{-\frac 12} ,
\qquad \rho = \sqrt{\det \theta^{-1}_{\mu\nu}} 
\label{sigma-rho-relation}
\eea
where
\be
g_{\mu\nu}(x) \, = \, \partial_\mu x^a\partial_\nu x^b\, \eta_{ab} 
 \,=  \,\eta_{\mu\nu} 
+  \partial_{\mu}\phi^i \partial_{\nu}\phi^j\delta_{ij} 
\label{g-def}
\ee
is the induced metric on $\cM$. 
$G_{\mu \nu}$ is dynamical, depending on 
the embedding fields $\phi^i$ and $\theta^{\mu\nu}$.
Standard embedding theorems imply that 
$G_{\mu\nu}$ can describe in
principle the most general metric in 4 dimensions.
Therefore the matrix model 
defines a theory of space-time and gravity
coupled to gauge fields and matter.

The matrix e.o.m. $[X^a,[X^b,X^{a'}]] \eta_{aa'} =0$ 
can be written in a covariant manner as \cite{Steinacker:2008ri} 
\be
\Delta_{G} \phi^i = \Delta_{G} x^\mu =0 .
\label{eom-varphi-0}
\ee
This implies the covariant equation  for
$\theta^{\mu\nu}$ 
\bea
G^{\g \eta}\, \nabla_\g (e^{\sigma} \theta^{-1}_{\eta\nu}) 
\, &=& \, e^{-\sigma}\,G_{\mu\nu}\,\theta^{\mu\g}\,\partial_\g\eta,
\label{eom-geom-covar-extra}\\
\eta(x) &=& \frac 14 e^\sigma\, G^{\mu\nu} g_{\mu\nu} 
\label{eta-def}
\eea
which relates $\theta^{\mu\nu}(x)$ with the metric $G^{\mu\nu}$.
Here $\nabla$ denotes the Levi-Civita connection with respect to
 $G_{\mu\nu}$.
In principle, $\theta^{\mu\nu} \neq 0$
 breaks (local) Lorentz
invariance. However, $\theta^{\mu\nu}$
does not enter explicitly the effective action 
to leading order, and its presence 
through higher-order terms
may  be below experimental limits
if the scale of noncommutativity $\Lambda_{NC}$ is 
high enough. In fact, this 
spontaneous breaking of Lorentz invariance 
leads to massless gravitons as discussed below. 
Note also that the matrix model defines 
preferred coordinates $x^\mu$, which are
not observable and not in conflict with observation. 

Equations \eq{eom-varphi-0} imply that
the embedding $\cM \subset \R^D$ is harmonic w.r.t.
$G_{\mu\nu}$.
A particularly interesting case
is given by geometries with
\be
G_{\mu\nu} = g_{\mu\nu} .
\label{g-G-relation-1}
\ee
It is not hard to see \cite{Steinacker:2008ya} that this holds
if and only if
$\eta = e^\sigma$,
which (in the Euclidean case) is equivalent to 
$\star \theta = \pm \theta$
where $\star$ is the Hodge star.
Then \eq{eom-geom-covar-extra} simplifies as 
\be
\nabla^\eta \theta^{-1}_{\eta \nu}  = 0 .
\label{eom-theta-3}
\ee 
These are formally the free Maxwell equations in the background
geometry $G_{\mu\nu}$. 
In particular, they  have propagating massless solutions
with 2 physical helicities. 
However, there are no charged fields under this 
``would-be'' $U(1)$ gauge field in the matrix model.
Rather, these modes turn into 
the 2 physical degrees of freedom of gravitons. 
Writing $\theta^{-1}_{\m\nu} = \bar\theta^{-1}_{\m\nu} + F_{\m\nu} $
on a flat (Moyal-Weyl) background, the metric fluctuations are
\be
h_{\mu\nu}  =  - \bar G_{\nu \nu'} \bar\theta^{\nu'\rho} F_{\rho\mu} 
- \bar G_{\mu \mu'} \bar\theta^{\mu'\rho} F_{\rho\nu} \, 
-  \bar G_{\mu\nu} F_{\rho\eta} \bar\theta^{\rho\eta}/2
\nn
\ee
which are nontrivial and Ricci-flat,
$R_{\mu\nu}[\bar G + h] = 0$ 
\cite{Rivelles:2002ez,Steinacker:2008ya}.

{\bf Quantization and induced gravity}: 
The quantization of the matrix model \eq{YM-action-extra} is defined by 
\be
Z = \int d X^a e^{-S_{YM}[X]}\, 
\ee
(omitting fermions for simplicity). Since one cannot simply
add an explicit Einstein-Hilbert term, the model
is highly predictive. The cosmological solutions given below 
add to the evidence that it
may provide a (near-?) realistic theory of gravity,
with great advantages  
for the cosmological constant problem.

Consider a perturbative quantization of the matrix model
around a given background as discussed above. Since all
fields couple to $G_{\mu\nu}$, standard considerations 
imply that in the classical-geometric limit,
the effective action at one-loop can be obtained from
Seeley-de Witt coefficients, so that
\be
\Gamma_{\rm 1-loop} = \frac 1{16\pi^2}\! \int d^4 x 
\sqrt{|G|}\,\left( c_1\Lambda_1^4 
+ c_4 R[G]\, \Lambda_4^2 + O(\ln \Lambda) \right) .
\label{S-oneloop-scalar}
\ee
The coefficients $c_i$ as well as the effective cutoffs $\Lambda_i$
depend on the detailed field content of the model, 
cf. \cite{Klammer:2008df}.
This is essentially the mechanism of induced gravity.

Now consider the equations of motion for the geometry,
taking into account the quantum contribution \eq{S-oneloop-scalar}.
Remarkably, 
equation \eq{eom-geom-covar-extra} for $\theta^{\mu\nu}$ is unchanged:
it is a direct consequence of Noether's theorem due to the  symmetry
$X^a \to X^a + c^a \one$, and therefore
protected from quantum corrections \cite{Steinacker:2008ya}.
However, the equation \eq{eom-varphi-0} for the embedding $\phi^i$ 
is modified at one loop.

Let us focus on the first term $\int d^4 x\sqrt{G}\, \Lambda^4$,
which is essentially the vacuum energy due to zero-point
fluctuations. In GR, it amounts to a huge 
contribution $\sim \L^4$ to the cosmological
constant, which must be canceled by an extremely fine-tuned
bare cosmological constant in order to 
reproduce the small value $O(meV)$ in the $\L$CDM model. 
This is the well-known cosmological constant problem, which persists even 
in models with $TeV$ scale supersymmetry.

We claim that this problem is resolved here.
To see this, note that
\be
|G_{\mu\nu}(x)| = |g_{\mu\nu}(x)| 
\label{G-g-4D}
\ee 
independent of $\theta^{\mu\nu}(x)$.
Hence the variation of the vacuum energy term
$$
\d \int d^4 x\sqrt{G} \, \sim \,
\int d^4 x\sqrt{g} g^{\mu\nu}\d g_{\mu\nu} \, \sim \, 
\int d^4 x\sqrt{g} \d \phi^i \Delta_g \phi^j \delta_{ij} 
$$ 
vanishes for harmonic embeddings 
\be
\Delta_g \phi^i=0 .
\label{harmonic-g}
\ee 
Therefore the term $\int d^4 x\sqrt{G}\, \Lambda^4$ has a different
physical meaning here:
It should not be interpreted as a cosmological constant,
but as a brane tension. In particular for harmonically embedded
branes (i.e. minimal surfaces), 
the precise value of its coefficient $\sim\Lambda^4$  is 
irrelevant and drops out from the equations of motion.
For example, flat Moyal-Weyl space is a solution even 
at one loop,
{\em without fine-tuning} $\Lambda$.
Thus minimally or harmonically embedded branes 
(w.r.t. $g_{\mu\nu}$)
are {\em protected from the  cosmological constant problem} 
\cite{Steinacker:2008ri}; they are
in fact stabilized through the vacuum energy. 
The crucial difference to general relativity is the 
parametrization of the geometry in terms of 
``tangential'' $\theta^{\mu\nu}$ and ``transversal'' 
$\phi^i$ rather than a
fundamental metric.  Gravitons originate from
fluctuations of $\theta^{\mu\nu}$
and are also blind to $\int d^4 x\sqrt{G}\, \Lambda^4$.

Combining \eq{eom-varphi-0} and \eq{harmonic-g}, we see that harmonic
embeddings with  $g_{\mu\nu} = G_{\mu\nu}$ 
solve the vacuum e.o.m. of the 
matrix-model {\em including quantum corrections}
\eq{S-oneloop-scalar}. The presence of matter 
will lead to deviations from harmonic embedding, however these
corrections are suppressed by factors $O(\frac{\rho}{\Lambda^4})$
where $\rho$ is the energy density. 
Thus $\Delta_G \phi^i = 0$ is valid
as long as the brane-tension $\sim \Lambda^4$
dominates the energy density of matter and curvature.

{\bf Robertson-Walker geometries}:
Now consider cosmological solutions of FRW type, and assume 
that $\Lambda_1 = O(TeV)$ to be specific.
The Friedmann equations
are replaced essentially by the 
requirement of minimal embedding  \eq{harmonic-g}
 with $g_{\mu\nu} = G_{\mu\nu}$.
This should be valid to high accuracy
except possibly in the very early universe
when the energy density becomes of order $O(TeV)$.
The FRW geometry has the general form
\be
ds^2 = -dt^2 + a(t)^2 d\Sigma^2
\label{FRW}
\ee
where
\be
d\Sigma^2 = \frac{1}{1-k r^2} dr^2 + r^2 d\Omega^2  
= d\chi^2 + S(\chi)^2 d\Omega^2 
\ee
is the 3-dimensional metric 
with uniform curvature;
here $S(\chi) = r = (\sin\chi,\chi,\sinh\chi)$ 
for $k=(1,0,-1)$ respectively.
The following provides harmonic embeddings of
FRW geometries for $k=\pm 1$ in $\R^{10}$, generalizing \cite{nielsen}:
\be
\vec x(t,\chi,\theta,\varphi) = \(\begin{array}{c}
\cR(t) \(\begin{array}{l} S(\chi)\sin\theta\cos\varphi \\
     S(\chi)\sin\theta\sin\varphi  \\
     S(\chi)\cos\theta \\
     C(\chi) \end{array}\)\\
0 \\  x_{c}(t)\end{array}\) \in \R^{10} \nn
\ee
where
\be
\cR(t) = a(t)\,\(\begin{array}{l} \cos\psi(t) \\
\sin\psi(t) \end{array}\) 
\ee
and $C(\chi) = (\cos\chi,\cosh\chi)$ for
$k=(1,-1)$ respectively. 
\begin{figure} [t]
\begin{center}
\includegraphics[scale=0.3]{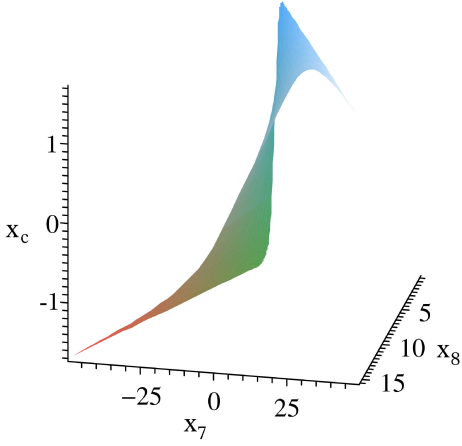}
\end{center}
\caption[dum]{\small{\label{fig:schlange} Embedding of universe in
    $\R^{10}$,
for $b=1,m=5$  }}
\end{figure}
Here $\eta_{ab}$ in \eq{YM-action-extra} is 
$\eta_{ab} =\diag(+,...,\pm,-)$ for
 $k=1$, or $\eta_{ab} =\diag(+,...+,-,-,+,+)$ for $k=-1$.
Then 
\bea
g_{\mu\nu}\, dx^\mu dx^\nu &=& 
\eta_{ab}\, d x^a d x^b  = 
- c(t) dt^2 + a(t)^2 d\Sigma^2
\label{FRW-embed-metric}, \nn\\
c(t) &\equiv& k(\dot{x_{c}}^2 - a^2 \dot\psi^2 - \dot a^2) .
\eea
This can be brought into FRW form \eq{FRW}
using a change of variables
$\frac{d\tau}{dt} =\sqrt{c(t)}$.
Equivalently, we can choose $t$ to be the ``proper'' time
variable $\tau$, such that $a(t)$ is the usual FRW scale parameter;
then $c(t) = 1$, and
\bea
\dot{x_c}^2 - a^2 \dot\psi^2 - \dot a^2 = k .
\label{c-constraint}
\eea
Note that the matrix coordinates and in particular
$x_c$ have no physical meaning from the brane point of view;
they are determined by the requirement
of harmonic embedding 
$\Delta_g x^a = 0$.
Due to the symmetry, it is enough to show that 
\bea
0 &=& \Delta_g (\cR(t) S(\chi)\cos\theta)  \nn\\
0 &=& \Delta_g x_c .
\label{xdot-eq}
\eea
This leads to
\bea
3\frac 1{a}\,(\dot a^2+k) + \ddot a - \dot\psi^2 a &=& 0 \label{addot-eq}\\
5 \dot \psi \dot a  + \ddot \psi a &=& 0 \label{psidot-eq}\\
3\frac 1{a}\,\dot a\dot x_c + \ddot x_c &=& 0 .
\label{a-eom}
\eea
These equations can be integrated as follows:
\bea
(\dot a^2+k)a^6 + b^2 a^{-2} &=& m = const  \nn\\
 \dot \psi  &=& b\, a^{-5} ,\qquad b = const >0  \nn\\
a^3 \dot{x_c} &=& d = const  ;
\eea
the last equation is in fact a consequence of 
\eq{c-constraint}, which gives $d= \sqrt{m}$ and hence $m>0$. 
This leads to
\bea
H^2 = \frac{\dot a^2}{a^2} &=& -b^2 a^{-10} + m a^{-8} -\frac {k}{a^2}.
\label{H-equation} \\
\frac{\ddot a}{a} &=& - 3 m a^{-8} + 4 b^2 a^{-10} .
\label{addot}
\eea
For small $a$ and $b\neq 0$, we find $\ddot a\gg 0$ 
which signals inflation. 
For large $a$, it follows that $\ddot a <0,\,\,\ddot a \to 0$. 
The case $b=0$ was obtained before in \cite{nielsen}.

For $k=0$, one finds an unrealistic age of the universe 
of order $4.5 \cdot 10^9$ years (assuming small $b$),
hence we will not pursue this case any further.
It turns out that $k=-1$ is the (near-) realistic 
case as discussed below.

To complete the solution of emergent gravity we need to find
a Poisson structure $\theta^{\mu\nu}$ 
which satisfies \eq{eom-theta-3} such that
$G_{\mu\nu} = g_{\mu\nu}$.
Clearly there exists no homogeneous and isotropic non-degenerate
$\theta^{\mu\nu}$.  However,  $e^{-\sigma}$ should at least 
be spatially homogeneous, because it 
determines the nonabelian gauge coupling
\cite{Steinacker:2008ya}.
Consider $k = - 1$. We introduce $\tilde t(t)$ through 
$\frac{dt}{a} = \frac {d\tilde t}{\tilde t}$,
and write the FRW metric in the form
\be
d s^2_g 
= \frac {a^2}{\tilde t^2} (-d\tau^2 + dr^2 + r^2 d\Omega^2) 
\ee
where $\tau = \tilde t \cosh(\chi), \,\, r = \tilde t \sinh(\chi)$.
In particular, for $a(t) = t$ we recover the well-known fact 
that the Milne universe is flat. 
Then the (complexified) symplectic form
\be
\theta^{-1} = \theta^{-1}_{\mu\nu} d x^\mu \wedge d x^\nu
= i d\tau \wedge dx_1 + dx^2 \wedge dx^3
\ee
in these flat coordinates 
is closed and 
($i$-) self-dual, hence \eq{eom-theta-3} is satisfied;
recall that this equation applies even at the quantum level.
Moreover we obtain $G_{\mu\nu} = g_{\mu\nu}$, and
\be
|g_{\mu\nu}| = \frac {a^8}{\tilde t^8}, \quad
|\theta^{-1}_{\mu\nu}| = 1, \quad
e^{-\sigma} = \frac {\tilde t^4}{a^4} .
\ee
In particular, $e^{-\sigma} \to 1$ for large $t$ as the geometry
approaches that of a Milne universe. Hence
the gauge coupling is approximately 
constant for large $t$, but
has a non-trivial time evolution in the early universe.

{\bf Emergent cosmology}:
The cases $k=0,+1$  imply a  too short
age of the universe given the present 
Hubble parameter. Therefore we focus on the case $k=-1$. 
Then $\dot a \to 1$ for large $a$, and the deceleration
parameter $q = - \frac{\ddot a a}{\dot a^2}\to 0$.
For large $t$, the time evolution approaches that of a Milne
universe, which is in remarkably good agreement with observation
 \cite{BenoitLevy:2008ia}. 
The age of the universe is found to be $\frac 1{H_0} \approx 13.9\cdot
10^9$ years, the time evolution of 
$a(t)$ in the $\Lambda$CDM model {\em at present}
being tangent with the evolution in the
Milne universe  \cite{Kutschera:2006bh}. 
Moreover, the main observational constraints
including the acoustic peak in the CMB background and the 
type Ia supernovae data appear to be consistent with 
an interpretation in terms of a Milne Universe \cite{BenoitLevy:2008ia}. 
While this geometry is excluded within GR, it 
makes perfect sense within emergent NC gravity.
In view of the unreasonable fine-tunings in the
presently favored $\Lambda$CDM model, 
this certainly deserves a more detailed investigation.

\begin{figure}[ht]
\vspace{0.5cm}
\begin{center}\hspace{-0.2cm}
\parbox{4cm}{\vspace{0cm}\includegraphics[scale=0.22]{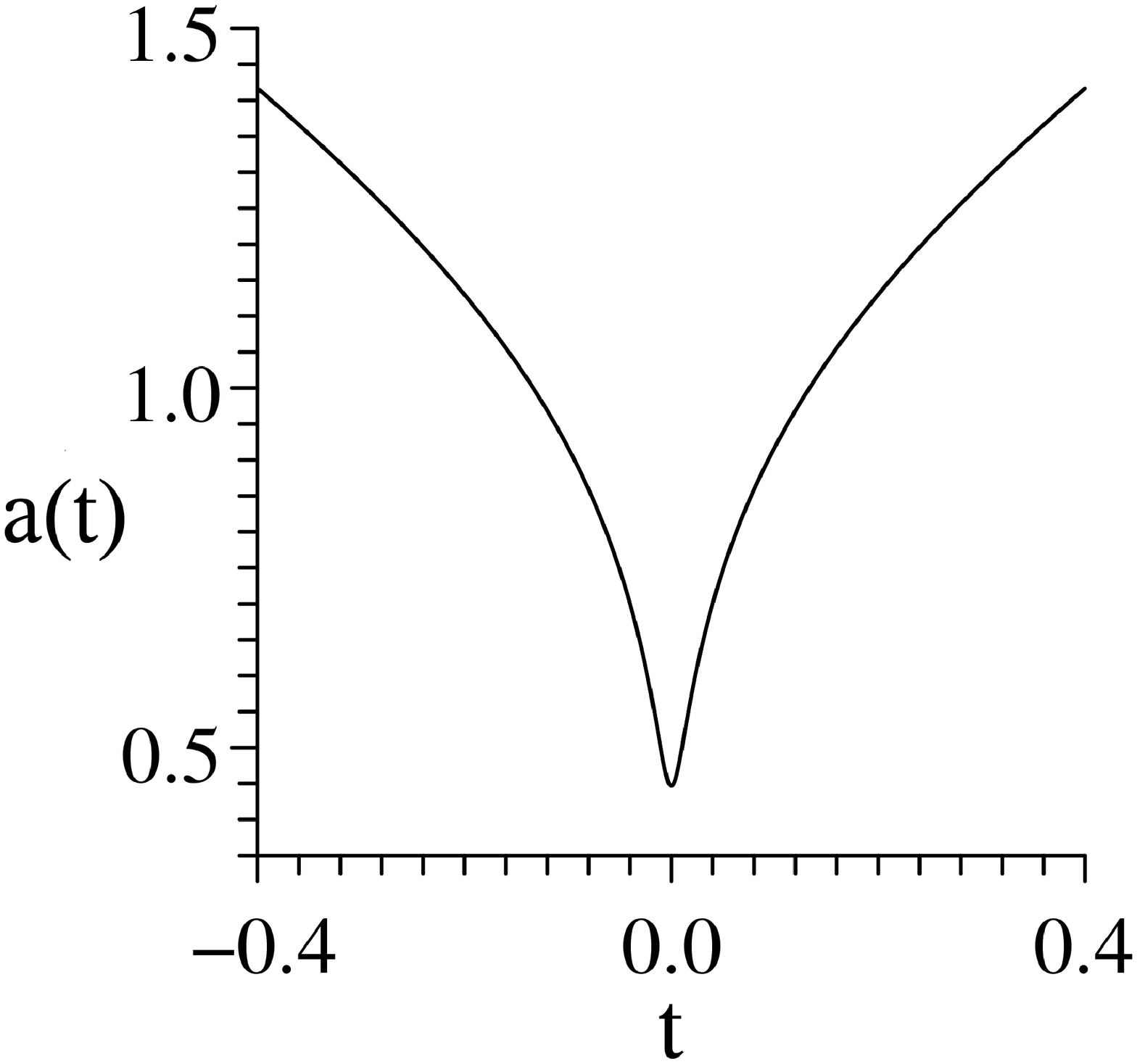}}
\hspace{0.2cm}
\parbox{4cm}{\vspace{0cm}\includegraphics[scale=0.22]{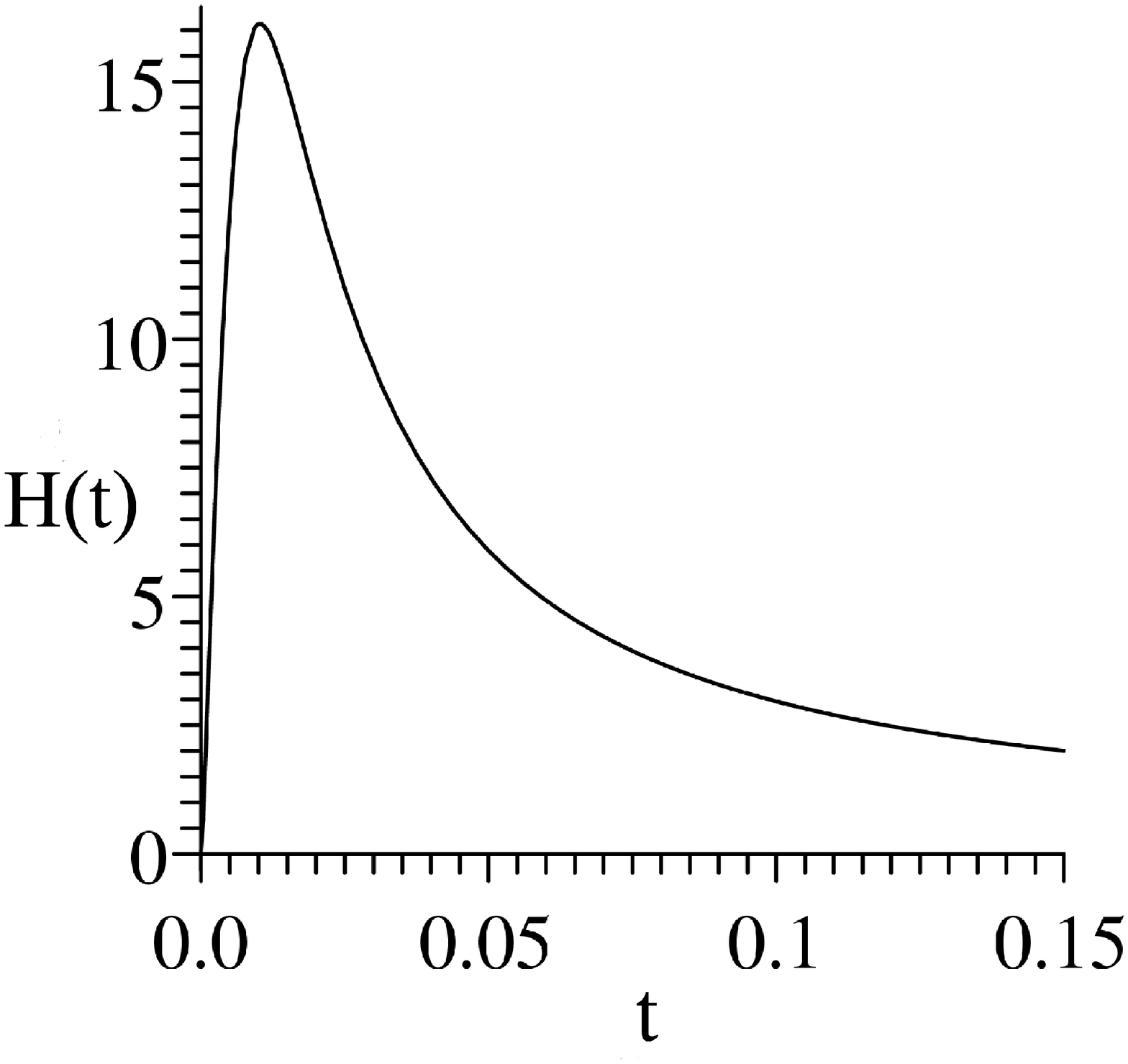}}
\end{center}
\vspace{-0.2cm}
\caption[dum]{\small{\label{fig:logbild} Evolution of 
$a(t)$ and
$H(t)$ for $m=5,\,b= 1$. }}
\vspace{-0.3cm}
\end{figure}

{\bf Inflation and big bounce}:
The scaling parameter is determined by 
\be
\dot a = \sqrt{-b^2 a^{-8} + m a^{-6} +1} .
\ee
For $b \neq 0$, denote with $a_0$ the (positive) root of the argument,
which is the minimal ``size'' of the universe. 
We fix the origin of time by $a(0) = a_0$, and define $t_1$ by
\be
\ddot{a}=0 \quad \Leftrightarrow \quad a(t_1) = \sqrt{\frac{4b^2}{3m}} = a_1
\ee
using \eq{addot}. 
Expanding $1+m\,a^{-6}-b^2\,a^{-8} = p (a-a_0) + ...$ 
around $a_0$ where 
$p = \frac{\mathrm{d}(1+m\,a^{-6}-b^2\,a^{-8})}{\mathrm{d}a}\vert_{a=a_0}$
shows an inflation-like phase 
$$
\dot{a}\sim\sqrt{p(a-a_0)}, \qquad
a(t)\sim\frac{p}{4}t^2 + a_0,
$$
which ends at $a(t_1)=\sqrt{\frac{4b^2}{3m}}$, where $\ddot{a}=0$
(``graceful exit''). A typical evolution of $a(t)$ 
 and the corresponding Hubble parameter is shown 
in figure \ref{fig:logbild}. 
Assuming $m^2 < b^3$, we have approximately  
$$
a_0 \sim b^{1/4}, \qquad
\frac{a(t_1)}{a_0}=\sqrt{\frac{4}{3}}\,\frac{b^{3/4}}{\sqrt{m}}.
$$
Note that the requirements for ``successful inflation'' such as a large
number of $e$-foldings will be greatly relaxed here compared with
standard cosmology; this should be addressed elsewhere.
For $b>0$, it is obvious that the time evolution should in fact not
start at $t=0$, but be completed symmetrically as
$$
a(-t) = a(t), \quad \psi(-t) = -\psi(t), \quad x_c(-t) = -x_c(t) 
$$
corresponding to a ``big bounce''  rather than a big bang.

We conclude that Yang-Mills matrix models
admit cosmological solutions which 
are in remarkably good agreement with observation. 
The type Ia supernovae data are accommodated without any fine-tuning
and without introducing dark energy.
The solutions should be valid
as long as the quantum-mechanical vacuum energy
dominates the energy density due to matter or radiation, i.e. 
up to epochs with TeV-range temperature
in the case of TeV-scale supersymmetry.
Some modifications could arise from compactification in
extra dimensions (leading to interesting low-energy gauge groups, 
effective scalar fields \& potentials as in \cite{Aschieri:2006uw}),
or soft SUSY breaking terms (e.g. a mass term) in the matrix model.
However it is unlikely that this would change our main conclusion, 
which is a Milne-like 
evolution after a big bounce and an inflation-like phase.
If confirmed this would resolve
the cosmological constant problem. 

{\bf Acknowledgments}
We want to thank H. Grosse, I. Sachs and H. Rumpf 
for useful discussions. 
This work was supported by FWF project P20017.

\end{document}